\documentstyle[prl,aps,epsf,preprint]{revtex}

\begin{document}

\tighten

\title{QCD Inequalities, Large \boldmath{$N_C$ and $\pi\pi$}\\ 
Scattering Lengths}
\author{S. Nussinov}
\bigskip

\address{School of Physics and Astronomy\\ 
Tel Aviv University, Tel Aviv, Israel\\
and\\
Department of Physics and Astronomy\\
University of South Carolina\\
Columbia, South Carolina 29208 USA\\
{~}}

\date{October 2000}

\maketitle

\begin{abstract}

In this short note we show that (I) in a QCD-like theory with four
(rather than two) degenerate flavors $ud\,u'd'$, the $\pi\pi'$ scattering
length is positive (attractive); and (II) in QCD with only two (u,d)
degenerate flavors the I=2 (say, $\pi^+\pi^+$ hadronic) scattering
length is, in the large $N_C$ limit, repulsive.  $\pi(\pi')$ are the
lowest physical states coupling to 
$J^p = \bar{u}(x)\gamma_5d(x)$ and $J^{p'} = \bar{u}'(x)\gamma_5d'(x),$ 
respectively.

\end{abstract}

\pacs{xxxxxx}

\narrowtext

To probe the threshold $\pi\pi$ (or ($\pi'\pi'$)) scattering consider
the Euclidean two-point correlators:
\begin{equation}
Q(x) = <0|J^p(x)J^p(x)(J^p(0)J^p(0))^+|0>
\end{equation}
\begin{equation}
Q'(x) = <0|J^p(x)J^{p'}(x)(J^p(0)J^{p'}(0))^+|0>
\end{equation}
where $J^p = \bar{u}(x)\gamma_5d(x)$ and $J^{p'} = \bar{u}'(x)\gamma_5d'(x),$ 
respectively.

The correlators have spectral decompositions
\begin{equation}
Q(x) = \int_{\mu_0}^{+\infty} \sigma_Q(\mu^2) {\rm exp}(-\mu x)d\mu^2
\end{equation}
\begin{equation}
Q'(x) = \int_{\mu_0}^{+\infty} \sigma_Q'(\mu^2) {\rm exp}(-\mu x)d\mu^2
\end{equation}
with a common threshold $\mu_0 = 2m_\pi$ (or $\mu_0 = m_\pi + m_{\pi'}$).
The asymptotic behavior of $Q(x)\, (Q'(x))$ in the limit $|x| \rightarrow
\infty$ (or Euclidear time separating the points $(0 = (\vec{0},0)$ and
$x = (\vec{0},t), \; t \rightarrow \infty)$ is controlled by the spectral density $\sigma_Q$ or
$\sigma_{Q'}$ at $\mu = 2m_\pi$.  The $\pi\pi$ (or $\pi\pi'$)
scattering at threshold can be described using nonrelativistic
methods.  The Levinson's theorem implies that the density of states in
any convenient large radius R of the $\pi\pi \, (\pi\pi')$ system
deviates from that of a free-noninteracting $\pi\pi \, (\pi\pi')$ pair
by
\begin{equation}
\frac{dn_{\pi\pi}}{d\mu^2} - \frac{dn_{\pi\pi}^{(0)}}{d\mu^2} = 
\frac{d\delta (k)^{(L=0)}}{dk} \approx a^{(L=0)}
\end{equation}
with $k = 1/2 \sqrt{\mu^2-4m_\pi^2}$ the CMS momentum, and 
$\delta^{(0)}$ and  $a^{(0)}$ are
the S-wave phase, shift and  scattering length of the $\pi\pi$ system.

Analogously for $\pi\pi'$,
\begin{equation}
\frac{dn_{\pi\pi'}}{d\mu^2} - \frac{dn_{\pi\pi'}^{(0)}}{d\mu^2} = 
\frac{d\delta (k)^{(L=0)}}{dk} \approx a'^{(L=0)} .
\end{equation}
Because hadronic forces have finite range only the $L=0$ wave needs to be 
considered. By definition,
\begin{equation}
\frac{dn_{\pi\pi}}{d\mu^2} \propto \sigma_Q(\mu^2); \;
\frac{dn_{\pi\pi'}}{d\mu^2} \propto \sigma_Q'(\mu^2) .
\end{equation}

To prove (I), we will show that
\begin{equation}
Q'(x) \geq \; <0|J^p(x)J^{p^+}(0)|0> <0|J^{p'}(x)J^{p'^+}(0)|0>.
\end{equation}
That is, the joint propagation of the four quarks---which
asymptotically becomes the joint $\pi$ and $\pi'$ propagation from 0
to $x$---is enhanced relative to the product of the separate $J^pJ^p$
and $J^{p'}J^{p'}$ correlators.  The latter is in fact the independent
$\pi$ and $\pi'$ propagations between 0 and $x$.

To prove Eq. (8), we use the path integral representation of the
various correlators.  Since all the four-quark flavors created at 0
and annihilated at $x$ are distinct, there is a unique ``contraction''
for $Q'(x)$ or Eq. (2) yielding:
\begin{eqnarray}
Q'(x) & = & \int d\mu(A)\;{\rm tr}\; \gamma_5S_{u_A}(x,0)\gamma_5S_{d_A}(0,x)
           \cdot {\rm tr}\; \gamma_5S_{u'_A}(x,0)\gamma_5S_{d'_A}(0,x)\\ 
\nonumber
      & = & \int d\mu(A) {\rm tr}(S_A^+(0,x)S_A(0,x)) \cdot 
           {\rm tr}\; (S_A^+(0,x)S_A(0,x))
\end{eqnarray}
where 
\begin{equation}
d\mu(A) = D(A_\mu(x)) \cdot e^{-SYM\{A_\mu(x)\}} \Pi_{u,d,u'd'} 
           Det(D\!\!\!\!/_A^{q_i}+m_i)/Z .
\end{equation}
is the positive\cite{Wei,VandW} measure in the path integral.
$D\!\!\!\!/_A = \gamma_\nu(\delta_\nu + gA_\mu^a\lambda_a)$ is the
covariant Dirac operator in the $A_\mu^a(x)$ background, and
$S_{i_A}(x) = <0|\frac{1}{D\!\!\!\!/_A+m_i^0}|x>$ is the propagator of
the Fermion (i.e., quark of flavor i and bare mass $m_i^0$) in the
$A_\mu^a(x)$ background.  Finally the partition function Z in the
denominator of Eq. (10) normalizes the path integral measure to $\int
d\mu(A) = 1$.

In deriving the second line of Eq. (9), we have used the ``$\gamma_5$
conjugation'' property of the Euclidean propagator,
\begin{equation}
\gamma_5 \; S_A^i(0,x)\gamma_5 = [S_A^i(x,0)]^\dagger
\end{equation}
with the $\dagger$ referring here to Hermitian conjugation in
color-spinor space of the ${S_A^i(x,0)}_{\alpha\alpha',aa'}\;\\ 
12~\times~12$ matrix with spinor $1 \leq \alpha, \alpha' \leq 4$ and
color $1 \leq a,a' \leq N_C$ indices, which for simplicity we generally
omit.  The latter $\gamma_5$ conjugation also implies the positivity
of the determinant of the Dirac operator $(D\!\!\!\!/_A+m^0)$, which is
the key to the claimed positivity of the measure $d\mu(A)$.  The
proofs of all these appear in the original paper of
Weingarten\cite{Wei} and of Vafa and Witten \cite{VandW} and in a
recent comprehensive review\cite{NandL}.  Finally, we utilized the
fact that since $m_u^{0} = m_d^{0} = m_{u'}^{0} = m_{d'}^{0}$, all
four propagators are actually the same.
\begin{equation}
S_A^u(x,0) = S_A^d(x,0) = S_A^{u'}(x,0) = S_A^{d'}(x,0) \equiv S_A(x,0).
\end{equation}
The inequality
\begin{equation}
Q'(x) \geq P(x) P'(x)
\end{equation}
with
\begin{equation}
P(x) = <0|J^p(x)(J^p(0))^+|0> \; ; \; P'(x) = <0|J^{p'}(x)(J^{p'}(0))^+|0>
\end{equation}
can be readily derived\cite{NandL}.  To this end we compare
$Q'(x)$---as given by the path integral (second line of Eq. (9))---and
the product of path integrals for $P(x),P'(x)$:
\begin{equation}
P(x) = \int d\mu(A)  {\rm tr}(S_A^+(0,x)S_A(0,x))
\end{equation}
and
\begin{equation}
P'(x) = \int d\mu(A)  {\rm tr}(S'^+_A(0,x) S'_A(0,x))
\end{equation}
(which in the $m_u^{(0)} = m_{u'}^{(0)};m_d^{(0)} = m_{d'}^{(0)}$ case of
interest are equal as $S_A = S'_A$).

Let us denote
\begin{equation}
{\rm tr}\; S_A^+(0,x) S_A(0,x) \equiv \pi_A(x) \geq 0 .
\end{equation}
The desired inequality (13) then is (as most QCD correlator
inequalities are) just the Schwartz inequality;
\begin{equation}
\int d\mu(A) \pi_A(x)\pi_A(x) \geq |\int d \mu(A) \pi_A(x)|^2.
\end{equation}
One dramatic way in which Eq. (18) could be implemented is if a
four-quark {\it bound} (scalar?) doubly-charged state
$\bar{u}\gamma_5d \; \bar{u'}\gamma_5d$ existed below the $\pi^+\pi'^+
(= 2m_\pi)$ threshold\cite{Espr}.  We are trying to mimic real QCD
where we know that exotic $q\bar{q}q\bar{q}$ mesons---should they
exist at all---are much heavier than $2m_\pi \approx$ 270 MeV.  Hence
we will {\it assume} that there are no such bound states below
threshold.

Using then Eqs. (5)-(7) above, our inequality (18) implies an
attractive (positive) S-wave $\pi\pi'$ scattering length; namely,
assertion I above.  One can also directly argue by going to the
nonrelativistic limit and viewing the path integral expression as an
euclidean diffusive evolution, that the probability of returning to
the origin $\vec{r}=0$ after some long time T is enhanced {\it iff}
the interaction between the $\pi\pi'$ is attractive.\cite{NusSat}  A related, more
ambitious approach uses four-point inequalities\cite{{NandSp},{Gupta}}.

One might argue that the introduction of $\pi'$ (and the corresponding
extra two flavors, $u'd'$, degenerate with $u$ and $d$) is unphysical
and renders this result meaningless.  Such an argument could be even
more forcefully made against Weingarten's proof\cite{Wei} of $m_N \geq
m_\pi$ where \underline{six} degenerate yet distinct flavors were
introduced.

We believe that this is not the case.  In particular, the
augmentation of the flavor sector can be useful for picking up
specific flavor contraction patterns in the real QCD with two (or
three) light (almost) degenerate flavors.  Thus, let us consider
$\pi\pi$ scattering in QCD (see Fig. 1).

\begin{figure}[tbhp]
  \begin{centering}
  \def\epsfsize#1#2{1.75#2}
  \hfil\epsfbox{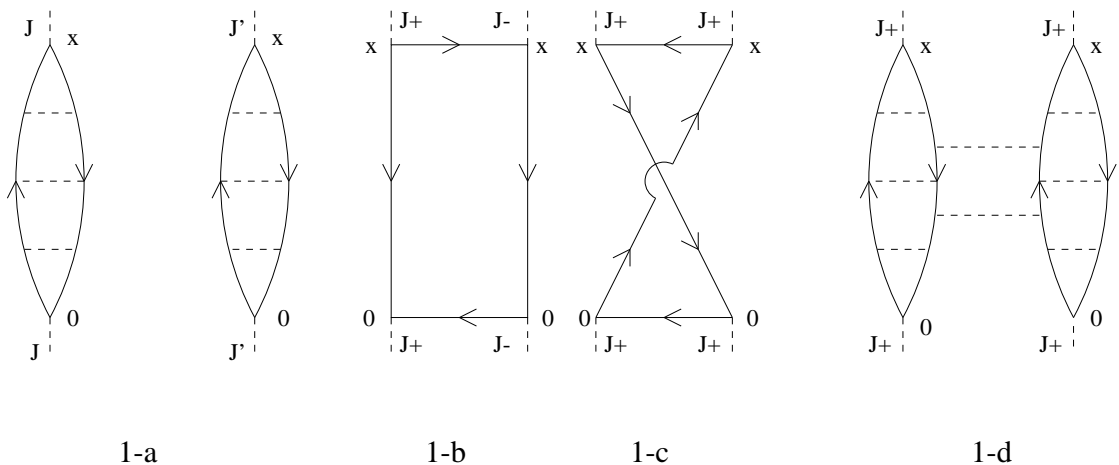}\hfil\hfill

\medskip

\caption{{\footnotesize Fig. 1a: Independent propagation of the $u\bar{d}$ and $u'd'$ blobs 
corresponding to the $\pi\pi'$ case.
Fig. 1b:  The anihilation contraction diagram for the case of 
$\pi^+\pi^-$.
Fig. 1c:  The exchange contraction diagram for the case of 
$\pi^+\pi^+$.
Fig. 1d:  The multi-gluon exchange diagram.}}

\label{ }
\end{centering}
\end{figure}

We can have separate, independent propagation of two pions,
represented in Fig. 1(a), where we have gluons exchanged only between
$u\bar{d}$ (or $\bar{u}d$) in the same initial and final pions. For
$\pi^+\pi^-$ scattering we have the flavor connected contraction
depicted in Fig. 1(b), corresponding to $q\bar{q}$ annihilation.
Likewise for $\pi^+\pi^+$ scattering, we have instead the quark exchange
diagram, Fig. 1(c).  Finally, for all pion pairs---and for the
$\pi\pi'$ case---we can have the $\bar{q}q$ bubbles interaction also
via gluon exchanges (at least two gluons are required because of the
color neutrality), as indicated in Fig. 1(d).

A key point is that for $\pi\pi'$ scattering we have \underline{only}
the free separate propagation (Fig. 1(a)) and gluon exchanges
(Fig. 1(d)): the $\pi$ and $\pi'$ have no common quarks and/or
anti-quarks to allow for $\bar{q}q'$ annihilation and/or $\dot{q}q'$
exchange.  

The same arguments could be made in any vectorial theory and, in
particular, in QED.  This then conforms to the attractive perturbative
two-photon exchange, $1/r^6$ van der Waals (VDW) potential and its
``retarded'' Casimir Polder version at large distances.  For two
polarizable atoms $A,B$ the latter is
\begin{equation}
V_{AB}^{CP} = - \frac{\hbar c}{(4\pi)^3r^7}
[23(\alpha^E_A \alpha^E_B + \alpha^M_A \alpha^M_B) 
-7(\alpha^E_A \alpha^M_B + \alpha^E_B \alpha^M_A)] 
\end{equation} 
The VDW interaction is, by second-order perturbation, always attractive
between two stationary systems in their ground state.  The CP
interaction is manifestly positive if A and B are dynamically the
same, i.e., 
$\alpha^E_A = \alpha^E_B$ and $\alpha^M_A = \alpha^M_B$.

Amusingly, the case for which we have been able to generalize this
result satisfies both requirements.  First, we know (and it can also
be proven via QCD inequalities) that the pseudoscalar $\pi$ (and
$\pi'$) are indeed the lightest mesons.  Second, since $m_u^0 = m_d^0
(= m_{u'}^0 = m_{d'}^0)$ and the gauge interaction are flavor
independent, the scattering ``A and B'', i.e., $\pi$ and $\pi$ (or
$\pi$ and $\pi'$) are dynamically the same.  We note that $V_{CP}$ of
Eq. (19) remains attractive so long as the objects A and B have
similar ratios of electrical and magnetical polarizabilities.  This is
expected to be the case for the ``chromo'' polarizabilities of
different hadrons, suggesting that $(J/\psi-A,Z)(D_s-A,Z)(D_s,B_u)$, 
etc., bound states of heavy/extended hadrons sharing no common quark 
flavors will form.

We next turn to our second main goal; namely, point II above.  With
only $u$ and $d$ degenerate quark flavors we have for $\pi^+\pi^-$
($\pi^+\pi^+$) scattering, or, for the correlators:
\begin{equation}
<0|J^{{p_s}^+}(x)J^{{p_s}^-}(x)(J^{{p_s}^+}(0)J^{{p_s}^-}(0)^+|0> 
= Q_{\pi^+\pi^-}
\end{equation}
\begin{equation}
<0|J^{{p_s}^+}(x)J^{{p_s}^+}(x)(J^{{p_s}^+}(0)J^{{p_s}^+}(0)^+|0> 
= Q_{\pi^+\pi}\; ,
\end{equation}
the contribution of the additional contractions.  These are
\begin{equation}
Q_{\pi^+\pi^-}(x)|_{Annihilation} = \int \, d\nu(A) {\rm tr} \, S_A^{+d}(0,x)S_A^u(x,x)S_A^d(x,0)S_A^{u^+}(0,0)
\end{equation}
and
\begin{equation}
Q_{\pi^+\pi^+}(x)|_{Exchange} = - \int \, d\nu(A) {\rm tr}\{S_A^+(0,x)S(0,x)S_A^+(0,x)S_A^d(x,0)S_A(0,x)\},
\end{equation}
respectively.
The crucial minus sign in Eq. (23) stems from the need to permute two
$\psi_u(x)$ (or two $\psi_u(0)$) noncommuting operators so as to
arrive from the $\int \, d\nu(A) {\rm tr}\{S_A^+(0,x)S_A^+(0,x)\}^2$
contraction pattern of Fig. 1(a) + 1(d) to Eq. (23) and the
contraction 1(c).

In the large $N_C$ limit the contributions (22) and (23) to the
correlators dominate.  It is tempting to associate these---in the
chiral, threshold limit---with the classical\cite{AandD} results,
$a^{I=0} \approx 0,2\,m_\pi^{-1}$ and $a^{I=2} = -\frac{2}{7} \;
a^{I=0}$.  Higher-order corrections in the chiral expansions preserve
the negataive (repulsive) $a^{I=2}$---in agreement with experimental
measurements which unfortunately are rather poor for $a^{I=2}$.  This
is particularly gratifying in view that we have an opposite sign
contribution (due to the multi-gluon, Fig. 1(d)) which does not, however,
reverse the sign.  Let us conclude with two comments.

{\it (i)}~The expressions (22) and (23) for $Q_{\pi^+\pi^-}$ and
 $Q_{\pi^+\pi^+}$ seem drastically different.  However, in the
threshold chiral limit they may both converge to similar objects.  The
point is that as $(x)$ (or $t$) tends to $\infty$, the relevant
propagating hadronic system is totally dominated by the threshold pion
states.  In the spirit of ``dual-string-QCD'' approach, which may not
be completely inappropriate\cite{Ros,Love} we may then represent the
soft pions as two nearby $u\bar{d}$ lines so that their mass
proportional to the string bit length between them approaches zero
and they are indeed point-like as appropriate for the chiral
Lagrangian approach.    Naively, we would then expect
the diagrams corresponding to Figs. 1(b) and 1(c) to be
essentially equal up to a factor (-1).  This would be the case if in
the VDM spirit we can describe the process merely via a $\rho$
exchange.  In this case, $a^{I=2} = -\frac{1}{2}\,a^{I=1}$.  To get a
better ratio we need some $I=0$ scalar $t$ channel exchange which
enhances $a_{\pi^+\pi^-}$ and suppresses $a_{\pi^+\pi^+}$.  The magnitude
of this term is fixed by requiring that it will have the same magnitude
also in the crosses, s, channel, thereby obtaining the desired
-2/7 ratio of the $I=2$ and $I=0$ scattering length.

{\it (ii)}~We can estimate the contribution of the multi-gluon
exchanges to threshold $\pi\pi$ physics by the positive shift of
$a^{I=0}$ and $a^{I=2}$ as compared with the original chiral
Lagrangian estimate.  It is tempting to identify the term,
\begin{equation}
4\pi a^2_{gluon\,exchange}
\end{equation}
with the asymptotic $\pi^+\pi^+$ cross section (controlled also by
multi-gluon exchange) extrapolates all the way to threshold.  We will
pursue this in a future publication.

\bigskip

The author would like to acknowledge the hospitality of the Department
of Physics of the University of Maryland.

\pagebreak


\begin{thebibliography}{99}
\bibitem{Wei} D. Weingarten, Phys. Rev. Lett. {\bf 51}, 1830 (1983).
\bibitem{VandW} C. Vafa and E. Witten, Nucl. Phys. {\bf B234}, 173 (1984).
\bibitem{NandL} S. Nussinov and M. A. Lampert, ``QCD Inequalities,'' 
hep-ph/9911532.
\bibitem{Espr} D. Espriu, M. Gross and J. F. Wehater, Phys. Lett. {\bf B146}, 
67 (1984).
\bibitem{NusSat} S. Nussinov and B. Sathiapalan, Nucl. Phys. {\bf B26}, 285 (1985.
\bibitem{NandSp} S. Nussinov and M. Spiegelglas, Phys. Lett. {\bf B202}, 265 (1988).
\bibitem{Gupta} R. Gupta, A. Patel and S. Sharpe, Phys. Rev. D {\bf 48}, 388 (1993).
\bibitem{AandD} S. Adler and R. Dashen, {\it Current algebra} (Benjamin, New York, 1968).
\bibitem{Ros} J. L. Rosner, Phys. Rev. Lett. {\bf 21}, 950 (1968).
\bibitem{Love} C. Lovelace, Phys. Lett. {\bf B28}, 264 (1968).


\end{thebibliography}
\end{document}